\documentstyle[12pt,aaspp]{article}
\begin{document}
\title{
GRAVITATIONAL LENSING EFFECT ON
COSMIC MICROWAVE BACKGROUND ANISOTROPIES: A POWER SPECTRUM APPROACH}

\author{Uro\v s Seljak\altaffilmark 1}
\affil{Department of Physics, MIT, Cambridge, MA 02139 USA}
\begin{quote}
\altaffilmark{}
\altaffiltext{1}{Also Department of Physics, University of Ljubljana,
Jadranska 19, 61000 Ljubljana, Slovenia}
\end{quote}

\def\bi#1{\hbox{\boldmath{$#1$}}}
\begin{abstract}
The effect of gravitational lensing on cosmic microwave background (CMB)
anisotropies is investigated using the power spectrum approach.
The lensing effect can be calculated in any cosmological model
by specifying the evolution of gravitational potential.
Previous work on this subject is generalized to a non-flat universe and to
a nonlinear evolution regime. Gravitational lensing cannot change the
gross distribution of CMB anisotropies, but it may redistribute the power
and smooth the sharp features in the CMB power spectrum.
The magnitude of this effect is estimated
using observational constraints on the power spectrum of gravitational
potential from galaxy and cluster surveys
and also using the limits on correlated ellipticities in distant galaxies.
For realistic CMB power spectra the effect on CMB multipole moments
is less then a few percent on degree angular scales, but gradually
increases towards smaller scales. On arcminute angular scales the
acoustic oscillation peaks may be partially or
completely smoothed out because of the gravitational lensing.

\end{abstract}
\newpage
\section{Introduction}

The effect of gravitational lensing on the cosmic microwave background
anisotropies has been studied in the past by several groups
(\cite{kashlinsky88}; \cite{blanchard};
\cite{cole};
\cite{sasaki}; \cite{tomita};
\cite{lindera},b;
Cay\' on, Mart\' inez-Gonz\' alez \& Sanz 1993a,b;
\cite{bassett}; Fukushige et al. 1994). Using
different approaches these authors came to very
different conclusions about the importance of the effect. Cole \&
Efstathiou (1989)
used a nonlinear CDM model
and found a small effect on CMB.
Cay\' on et al. (1993a) used a linear model and found
an appreciable effect on
arcminute angular scales for some models.
On degree angular scales they also found a negligible effect.
On the other hand, using different approaches
such as the Dyer-Roeder distance-redshift relation or simplified N-body
simulations
Bassett et al. (1994) and Fukushige et al. (1994)
found a significant effect even on
degree angular scales.

There are several shortcomings of these studies
that do not allow one to draw a firm conclusion
on the importance of the lensing effect on CMB.
First,
the studies are based on a particular cosmological model
and the results could change significantly if the model is changed.
While some groups (e.g. \cite{cole}; \cite{cay93a})
attempted to asses this uncertainty by presenting results
for different viable cosmological models, others (e.g. \cite{bassett},
Fukushige et al. 1994) used models that do not allow a direct comparison with
existing observational constraints and thus may not even be realistic models.
A second shortcoming of previous studies is that they do not fully
include the evolution of large-scale structure in their models.
While Cole \& Efstathiou (1989) calculated the effect only at late epochs
when the matter is in the
nonlinear regime, Cay\' on et al. (1993a,b) only included the linear
evolution, whereas
Bassett et al. (1994) and \cite{fuk} neglected any evolution at all
and assumed that the universe did not change from a certain
redshift until today.

The purpose of this paper is to provide a more realistic answer on the
importance of the effect by using observational constraints on large-scale
structure distribution and properly including its evolution.
The method used is based on the power spectrum approach in linearized
gravity and is
similar to the one used by Linder (1990a,b), Kaiser (1992) and Cay\' on et
al. (1993a,b).
An equivalent method based on optical scalars
has been developed by Gunn (1967) and extended by Blandford \& Jaroszynski
(1981). Present work differs from previous studies in
that I also include the nonlinear effects by modeling the power
spectrum evolution in the nonlinear regime.
By comparing the nonlinear calculation to the linear approximation one
can identify the angular scale where the nonlinear effects become important.
I also extend the calculation to the
case of an open (or closed) universe and correct some erroneous
expressions in the literature, all of which allows to calculate
the lensing effect in any standard cosmological model
(i.e. in any model
based on a weakly perturbed metric in a
universe that is homogeneus and isotropic on large scales).
The estimate of the lensing effect on the CMB is based on the
observational constraints on the power spectrum and on
the ellipticity correlations of distant galaxies,
which enables to asses its magnitude in our universe.
The results are presented in terms of the CMB anisotropy power spectrum, which
allows one to discuss the effect independent of the observational strategy.
In the conclusion section I
discuss the possible sources of discrepancy between present results and
some of the previous work on this subject.

\section{Formalism}
In this section I review the formalism to compute the
gravitational lens effect on a pair of propagating photons
separated by an angle $\theta$ at the observer's position.
The starting point is a perturbed Robertson-Walker
model with small-amplitude scalar metric fluctuations. In the
longitudinal gauge
(\cite{bard80}) one can write the line element as
(adopting c=1)
\begin{equation}
ds^2=a^2(\tau)\biggl[-(1+2\phi)d\tau^2+(1-2\phi)
[d\chi^2+\sin_K^2\chi(d\theta^2+\sin^2 \theta d\phi^2)]
\biggl]\ ,
\label{metric}
\end{equation}
where the metric is expressed with comoving spherical coordinates and
conformal time $\tau$ and $a(\tau)$ denotes the expansion factor.
I defined
\begin{eqnarray}
\sin_K\chi \equiv
\left\{ \begin{array}{ll} K^{-1/2}\sin K^{1/2}\chi,\ K>0\\
\chi, \ K=0\\
(-K)^{-1/2}\sinh (-K)^{1/2}\chi,\ K<0\\
\end{array}
\right.
\end{eqnarray}
Curvature $K$ can be expressed using the present density
parameter $\Omega_0$ and the present
Hubble parameter $H_0$ as $K=(\Omega_0-1)H_0^2$.
The
total density $\Omega(\tau)$ is in general time dependent and
can have contributions from mass density
$\Omega_m(\tau)$ or
vacuum energy density $\Omega_{v}(\tau)$, $\Omega(\tau)=\Omega_m(\tau)+
\Omega_v(\tau)$. The metric perturbation
$\phi$
can be interpreted as the Newtonian
potential since, neglecting the contributions from
wavelengths larger than the Hubble distance,
it obeys the cosmological Poisson equation
\begin{equation}
\nabla^2\phi={3 \over 2}H^2\Omega_m(\tau)a^2\delta  ,
\end{equation}
where $\delta$ is the mass density fluctuation.
Statistical
properties of the potential on scales small compared to the curvature scale
can be described with its Fourier transform
$\phi(\vec k,\tau)$, where $
\phi(\vec x,\tau)=\int d^3k \phi(\vec k,\tau)e^{i\vec k \cdot \vec x}$.
Its ensemble mean and variance
are $\langle \phi(\vec k,\tau)\rangle=0$
and $\langle \phi(\vec k,\tau)\phi^*(\vec k',\tau)\rangle=
P_{\phi}(k,\tau)\delta^3(\vec
k-\vec k')$, where $P_{\phi}(k,\tau)$ is the power spectrum of the potential
at time $\tau$.

A photon propagating
through the universe will be deflected
by the mass concentrations along its path.
The rate of change in the photon direction $\vec n$ is given by
the photon geodesic equation, which applied to the metric in equation
\ref{metric} gives
\begin{equation}
{d\vec n \over dl}=2\vec n \times (\vec n \times \vec \nabla
\phi)\equiv -2 \vec \nabla_\perp \phi,
\label{dndl}
\end{equation}
where the symbol $\vec \nabla_\perp \phi$ denotes the transverse
derivative of the potential and
$l$ is the comoving path length along the photon geodesic.
Gravitational potential $\phi$
can be viewed as providing a force deflecting the photons
while they
propagate through the
unperturbed space-time, described by a 3-sphere (closed
universe), 3-hyperboloid (open universe)
or Euclidean space (flat
universe).
Because the only observable photon direction is that at the observer's
position it is convenient to propagate photons relative to their final
direction
(i.e. backwards in time).
Gravitational lensing is not expected to lead to large deflection angles
(e.g. Linder 1990; Seljak 1994)
and one can
replace the transverse derivatives in equation (\ref{dndl})
with the transverse derivatives with respect to the observed direction
of the photon or with respect
to any other direction that has a small angular
separation with the photon. In this spherical plane approximation
the observed direction of the photon can be described with a two-dimensional
angle $\vec \theta$ with respect to the origin. Moreover, the null geodesic
condition for photons gives $d\tau \approx -d\chi$ neglecting corrections of
the order O($\phi$). Even when metric perturbations are present, one can
continue to parametrize the geodesic with the unperturbed comoving
radial distance $\chi$ or the conformal time $\tau$, which are related
through $\chi=\tau_0-\tau$, where $\tau_0$ is the conformal
time today.

The total deflection angle between
the photon source at the last-scattering surface\footnote[1]{I assume
throughout
the paper that CMB anisotropies are generated at a recombination time
$\tau_{\rm rec} \approx 0$ and therefore $\chi_{\rm rec} \approx \tau_0$.
This is a good approximation for all the models
where the CMB fluctuations are generated at a high redshift.}
and the observer
is given by
\begin{equation}
\delta \vec \alpha=-2\int_0^{\chi_{\rm rec}}\vec \nabla_\perp\phi d\chi.
\end{equation}
Similarly, the photon angular
excursion on the last-scattering surface relative to its observed value
is given by
\begin{eqnarray}
\delta \vec \theta=-2\int_0^{\chi_{\rm rec}}
W(\chi, \chi_{\rm rec})\vec \nabla_\perp\phi d\chi, \nonumber \\
W(\chi, \chi_{\rm rec})={\sin_K(\chi_{\rm rec}-\chi)\over
\sin_K\chi_{\rm rec}}.
\end{eqnarray}
In a flat universe the latter simplifies to
$W(\chi,\chi_{\rm rec} )=(1-\chi/\chi_{\rm rec})$.
Note that it is $\delta \vec \theta$ that is
relevant for the discussion of lensing effects on CMB, because one is
interested in the angular excursion of a photon on the CMB last-scattering
surface and not in the change in its direction.
Some of the previous work on this subject used
$\delta \vec \alpha$ instead of $\delta \vec \theta$ (e.g.
\cite{cay93a}).
As shown by Seljak (1994), in a flat $\Omega_{m0}=1$
linear theory this leads to a
factor of $10^{1/2}$ overestimate of the relative dispersion between two
photons.
In the following I will restrict the discussion to
$\delta \vec \theta$.

Two photons $A$ and $B$ observed with an angular separation
$\theta$ have a different angular separation when emitted from the
source position. Its mean is equal to the unperturbed value,
while  the dispersion is given by (Seljak 1994),
\begin{eqnarray}
\sigma({\theta})&=&2^{-1/2}\biggl\langle \biggl[
 \delta\vec \theta^A-
\delta\vec \theta^B\biggl]^2
\biggl\rangle_\theta^{1/2}=
\biggl[C_{{\rm gl}}(0)-C_{{\rm gl}}(\theta)\biggl]^{1/2} \nonumber \\
C_{{\rm gl}}(\theta )&=&
16\pi^2\int_0^{\infty}k^3dk\int_0^{\chi_{\rm rec}}P_{\phi}(k,
\tau=\tau_0-\chi) W^2(\chi, \chi_{\rm rec}) J_0(k\theta\sin_K\chi)d\chi ,
\label{limber}
\end{eqnarray}
where $\langle \rangle_\theta$ denotes the ensemble average performed over
all pairs of photons with a fixed observed angular separation $\theta$ and
$J_0(x)$ is the Bessel function of order 0.
The derivation of equation \ref{limber} is based on Limber's equation
in Fourier space
(e.g. \cite{kaiser}), which
assumes that the dominant scales contributing to
the dispersion are much smaller than the photon
travel distance. This condition is
satisfied
for sources at cosmological distances. No assumption on the
power spectrum has been made and equation \ref{limber}
can be used both in the linear and in the non-linear regime, both of which can
be
described by a time evolution of the power spectrum $P_{\phi}(k,
\tau)$. I neglected the Poisson contribution arising from
the discrete nature of galaxies, which is only important if $\Omega_{m0}<0.1$
(\cite{blandford}; \cite{cole}).
An important approximation needed to derive equation \ref{limber} is that
the potential sampled
by the perturbed photon geodesic can be replaced with the potential
along the unperturbed geodesic. This limits its applicability to the
regime where $\sigma(\theta)/ \theta\ll 1$.
When $\sigma(\theta)/ \theta $ becomes of order unity
the above assumption is no longer valid and the
separation between the photons may grow significantly faster than what the
model would predict. This point will be discussed again in the last section.

It is useful to give a physical understanding of the lensing effect on
two nearby photons. For simplicity I restrict the discussion to
scattering on a single scale
$k^{-1}$ and to a flat space.
In real space $k^{-1}$ is a correlation length and
determines the scale on which regions become uncorrelated. For sufficiently
small angles the photons
separated by distance $\chi\theta$
propagating through a region of size $k^{-1}$ are coherently
deflected (assuming $\chi\theta<k^{-1}$).
The change in the relative angle between the two after crossing
this region is given by
$\delta \theta_1 \approx \chi \theta \nabla_\perp (2 k^{-1}\nabla_\perp \phi)
\approx 2k\chi \phi \theta$.
The photons pass through $N \approx
k\chi_{\rm rec}$ uncorrelated regions
and the total rms
deflection angle between the photons grows in a random walk fashion,
$\delta \theta=N^{1/2}\delta \theta_1$ or $\sigma(\theta)/\theta \propto
(k\chi_{\rm rec})^{3/2}\phi$.
Adding the contributions from different
modes one reproduces, numerical factors aside, the small angle limit of
equation \ref{limber}. For large separation angles
the scattering is incoherent ($\chi\theta>k^{-1}$)
and each photon is deflected by $(k\chi_{\rm rec})^{1/2}\phi$,
independently of $\theta$,
implying $\sigma(\theta)/\theta \to 0$. This asymptotic behavior is
confirmed by the numerical results presented in the next section.

Once $\sigma( \theta)$ is known as a function of $\theta$ it is
straightforward to calculate the lensing effect on the
CMB fluctuations. The effect is most easily expressed in terms of
the temperature anisotropy
correlation function $C(\theta)=\langle {\Delta T \over T}^A
{\Delta T \over T}^B\rangle_\theta$.
Using the two-dimensional formalism and isotropic approximation
presented in the Appendix
one obtains the modified correlation function
$\tilde{C}(\theta)$,
\begin{equation}
\tilde{C}(\theta)={1\over \sigma^2(\theta)}\int_0^{\pi}\beta d\beta C(\beta)
e^{-(\beta^2+\theta^2)/2\sigma^2(\theta)}I_0\left[{\theta \beta \over
\sigma^2(\theta)}\right],
\label{c}
\end{equation}
where $I_0$ is the modified Bessel function of order 0. This equation
is strictly valid only for gaussian fluctuations,
but should give a reasonable
estimate of the effect even when this condition is not satisfied.
Note that the effect of lensing is to integrate the correlation function with
approximately a gaussian centered at $\theta$ with dispersion $\sigma(\theta)$,
as can be seen using the asymptotic expansion of $I_0$ combined with the
exponential in equation \ref{c}. Thus,
lensing acts as a filter
smoothing out the sharp features in the correlation function.
For lensing to be important the correlation function at $\theta$
must be changing rapidly on a scale $\sigma(\theta)$.

It is customary
to present different models of CMB anisotropies
in terms of the power spectrum, given
by the multipole
moments $C_l$. These are obtained by the Legendre transform of correlation
function, $C_l=2\pi\int_0^\pi \sin \theta C(\theta)P_l(\cos \theta)d\theta
$, where $P_l(x)$ is the Legendre polynomial of order $l$.
The lensing effect on the $C_l$ multipoles can be efficiently
calculated using the
Gauss-Legendre integration of $\tilde{C}(\theta)$ in equation
\ref{cthclful}.
One can estimate the effect on $C_l$
by assuming $\epsilon=\sigma(\theta)/\theta$ is a constant (Bond 1995).
While this is not true
in general (figure \ref{fig1}), one may hope to use this approximation
if $\sigma(\theta)/\theta$ is sufficiently slowly changing with
$\theta$ and is small, so that only correlations over a narrow range
of $\theta$ are mixed by lensing.
In this case $\epsilon$ should be determined by
its value at $\theta \propto l^{-1}$ (\cite{Bond95}). From equation
\ref{cleps} follows in the $\epsilon \ll 1$ limit
\begin{equation}
\tilde{C}_l=\int_0^\infty C_{l'}{dl' \over \sqrt{2 \pi}\epsilon l'}
e^{-(l-l')^2/2(\epsilon l')^2}.
\label{clsmooth}
\end{equation}
Lensing thus smoothes the spectrum of $C_l$ with a gaussian of relative width
$\epsilon$,
similar to the effect on the correlation function.

One way to calculate $\sigma(\theta)$
is to use the observational constraints on the power
spectrum from the large-scale structure observations,
carefully including the effects of
the evolution in a given cosmological model. This approach will be
explored in the next section.
A somewhat less model-dependent estimate can be obtained from
the observational constraints
on correlated distortions of distant galaxy images.
This can be described by
$p(\theta)$,
the average polarization
within a circular aperture of radius $\theta$,
which describes the correlations in the ellipticities of galaxy
images as a function
of angle $\theta$.
It is
related to the power spectrum using a $\Omega \ne 1$
generalization of the expression
given by
Blandford et al. (1991) and Kaiser (1992),
\begin{equation}
p^2(\theta)=16\pi^2
\int_0^{\infty}k^5dk\int_0^{\chi_{\rm g}}P_{\phi}[k,\tau=\tau_0-\chi]
W^2(\chi,\chi_{\rm g})\sin_K^2\chi
\left[{2J_1(k\theta\sin_K\chi) \over k\theta\sin_K\chi}\right]^2d\chi,
\label{p}
\end{equation}
where I assumed for simplicity that all the galaxies lie at the same
source position $\chi_{\rm g}$.
Using a small argument
Taylor expansion of Bessel functions in equations \ref{limber} and
\ref{p} one obtains a simple scaling between
$\sigma(\theta)$ and $p(\theta)$ for a flat
$\Omega_{m0}=1$ universe in the linear
regime, independent of the power spectrum on small angular scales,
$\sigma( \theta)/ \theta=
2^{-1}p(\theta)
(\chi_{\rm rec}/\chi_{\rm g})^{3/2}$.
Nonlinear effects and $\Omega_{m0}<1$ make the low redshift contributions more
important relative to the case above, which decreases $\sigma( \theta)/
\theta$ derived from $p(\theta)$. Numerical evaluation confirms this
prediction
and so the scaling above can be used
to give an upper limit on $\sigma( \theta)/\theta$ from the observational
limits on $p(\theta)$ on arcminute scales.

\section{Estimate of the Lensing Effect in Our Universe}

To compute the lensing effect one needs to specify the power
spectrum of potential as a function of scale and time. In linear
regime the time dependence of density perturbation in a CDM
dominated universe obeys the well known growing mode solution. For
the particular case of $\Omega_{m0}=1$ universe the potential does not
change in time and lensing contributions at early times are as
important as those at late times.
For the nonlinear evolution of the power spectrum one can either use
N-body simulation results or adopt a semianalytic approximation for it.
Here I adopted the approximation given by
\cite{hamilton}, generalized to $\Omega \ne 1$ by \cite{peacock} and to
the density power spectra with slopes $n<-1$ by \cite{bhuv}.
This prescription is based on an educated guess of what the
evolution of the density correlation function should be in the nonlinear
regime.
Although not exact, it agrees well with the
results of N-body simulations (see Peacock \& Dodds 1994 and Mo et al. 1995
for a detailed discussion of its applicability)
and should give a good estimate of the
nonlinear power spectrum in the regime where dissipative baryonic
processes can be neglected.
The linear to nonlinear mapping is most easily
expressed using the  mass density variance $\Delta^2(k)$, which is
related to the potential power spectrum via
$\Delta^2(k)=16\pi k^7P_\phi(k)/9\Omega_m(\tau)^2H^4a^4$.
The relation between the linear and nonlinear power spectrum is given
by
\begin{eqnarray}
\Delta^2(k_{nl})&=&f_{nl}[\Delta^2(k_l)]; \ \ \
k_l=[1+\Delta^2(k_{nl})]^{-1/3}k_{nl} \nonumber \\
f_{nl}(x)&=&x\left[ { 1+0.2\beta x+ (Ax)^{\alpha \beta}
\over 1 + ([Ax]^\alpha
g^3(\Omega_m,\Omega_v,a)/[11.68x^{1/2}])^\beta}\right]^{1/\beta},
\label{fnl}
\end{eqnarray}
where $A=0.84[g(\Omega_m,\Omega_v,a)]^{0.2}, \alpha=2/g(\Omega_m,\Omega_v,a)$
and
$\beta=2g(\Omega_m,\Omega_v,a)$. The linear growth factor
$g(\Omega_m,\Omega_v,a)$
can be approximated
with a few percent accuracy as (\cite{lahav}; \cite{carroll})
\begin{eqnarray}
g(\Omega_m,\Omega_v,a)&\approx & {5\Omega_{m0}\over 2[
X(1+[\Omega_{m0}/aX]^{0.6})-a^2\Omega_{v0}+\Omega_{m0}/2a]} \nonumber \\
X&=&1+\Omega_{m0}(a^{-1}-1)+\Omega_{v0}(a^2-1).
\end{eqnarray}
Mapping in equation \ref{fnl} can be improved by allowing for the
variation in the shape of the power spectrum.
One can introduce an effective
index of the density power spectrum
$n_{\rm eff}=d\ln[k^4P_\phi(k)]/d\ln k$
at a wavevector $k$ defined such that the rms mass fluctuations averaged
within a
sphere of size $k^{-1}$ is unity ($\sigma_{k^{-1}}=1$).
For $n_{\rm eff}=0$ the
mapping above gives reliable results, while for $n_{\rm eff}<-1$
there are substantial
deviations from the N-body simulations. As shown by Mo et al. (1995),
one can improve this by replacing
$\Delta^2$ by $\Delta^2/B(n_{\rm eff})$, where
\begin{equation}
B(n)=0.795\left[{\Gamma\left({17+n \over 10+2n}\right) \over
\Gamma \left({11+n\over 10+2n}\right)}\right]^{-(5+n)}.
\end{equation}
In this case a better fitting formula for the $\Omega_{m0}=1$ case is
given by (Mo et al. 1995)
\begin{equation}
f_{nl}(x)=x \left({1+2x^2-0.6x^3-1.5x^{7/2}+1x^4 \over 1+0.0037x^3}
\right)^{1/2}.
\end{equation}

The system of equations presented above
can be used to calculate $\sigma(\theta)/\theta$ for most cosmological
models of current interest (one exception being the models with
massive neutrinos on small scales where neutrino free streaming is
important).
To obtain an estimate of the lensing effect in our universe
I will use observational constraints on the power spectrum, as
compiled by \cite{peacock}.
The
power spectrum can be parametrized with a CDM type linear transfer function
(\cite{BBKS}) with two free parameters, the amplitude
$\sigma_8$, determined by the mass fluctuation averaged within a
sphere of radius $8h^{-1}$Mpc
and the shape parameter $\Omega_{m0} h$, determined by the
turnover position in the power spectrum
($h$ is the present day Hubble parameter in units of 100km/s/Mpc).
For wavevectors between $10^{-2}$ and 1 $h$Mpc$^{-1}$
all the galaxy and cluster surveys are in a reasonable agreement
with a CDM type linear power spectrum with $\Omega_{m0} h\approx 0.25$
(Peacock \& Dodds 1994; \cite{cfa}).
For normalization I will adopt
$\sigma_8=0.8$, which
is close to the normalization obtained by \cite{peacock} and by
\cite{WEF} using
the cluster abundances normalization over most of the interesting
range of $\Omega_{m0}$.
This normalization also agrees with the COBE normalization for
the currentlty favored $\Omega_{m0}=0.4$, $h=0.65$ case,
both in the open universe model
(\cite{gorski95}) and in the cosmological constant dominated model
with a modest tilt (Stompor, G\' orski \& Banday 1995).
The adopted power spectrum is likely to be within
a factor of two of the real power spectrum
on the arcminute scales and larger.

\begin{figure}[t]
\vspace*{7.3 cm}
\caption{$\sigma(\theta)/\theta$ versus $\theta$ for 3 different values of
$\Omega_{m0}$ and $\Omega_{v0}$ using the power spectrum
with $\Omega_{m0}=0.25$ and
$\sigma_8=0.8$.
Thick lines are the result of a full nonlinear
calculation, while the thin lines give the corresponding linear case.
Also indicated are the 90\% c.l. upper limits
from ellipticity correlations of distant galaxies, as derived from
observations by Fahlman et al. (1994) (A) and Mould et al. (1994) (B).}
\includegraphics{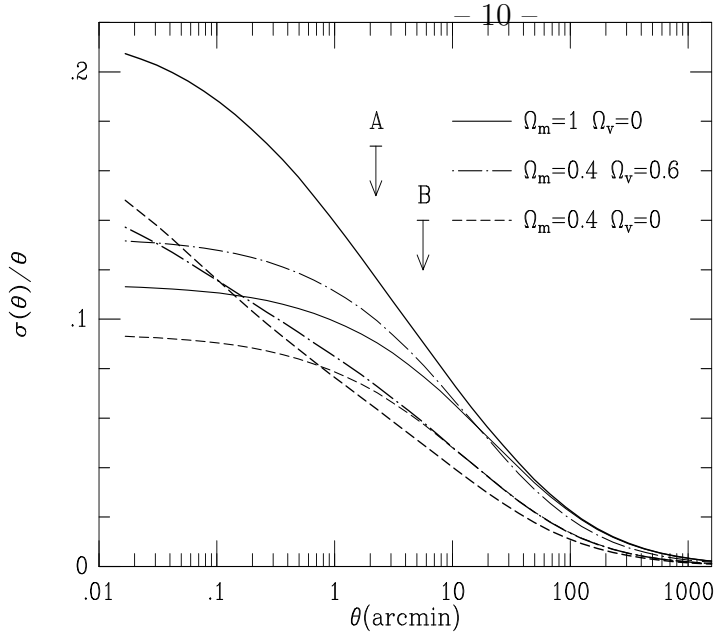}

\label{fig1}
\end{figure}

Given the linear power spectrum and its nonlinear evolution one can
compute $\epsilon=\sigma(\theta)/\theta$ as a function of $\theta$.
In figure \ref{fig1} the results are
presented for the power spectrum discussed above in flat
$\Omega_{m0}=1$,
flat low $\Omega_{m0}$ model and open low $\Omega_{m0}$ model, all of
which are known to phenomenologically agree with most of the large-scale
structure observations.
The thick curves give the result of a full nonlinear calculation, while
the thin curves show the corresponding linear case.
One can see that
while in the linear case $\sigma(\theta)/\theta$ approaches to a constant
for small $\theta$, it continues to increase in the nonlinear case.
Therefore in the real universe one cannot define a typical
coherence angle, which was used by previous studies (e.g.
\cite{sasaki}; \cite{lindera}; \cite{cay93a}) and the approximation
$\epsilon={\rm const}$ is not valid on any scale.
The results can only be used on angular scales above a few arcseconds,
where $\sigma(\theta)/\theta \ll 1$
and where the nonlinear mapping (based on the evolution of
collisionless matter) gives reliable estimates.

As seen in figure \ref{fig1} in the linear regime the lensing effect on CMB
decreases with $\Omega_{m0}$. This is mainly due to the
linear decrease of potential with $\Omega_m$ in the Poisson equation,
partly offset by the longer travel distance,
larger growth factor ratio
$g(\Omega_m,\Omega_v,a)/g(\Omega_{m0},\Omega_{v0},a_0)$ and in the
nonlinear regime by
larger nonlinear effects in the low $\Omega_{m0}$ models.
The latter is more important
because the scales that are nonlinear today became nonlinear earlier than
the corresponding scales in an $\Omega_{m0}=1$ universe.
In a low $\Omega_{m0}$ model
the universe changes from $\Omega_{m0} \approx 1$ to $\Omega_{m0} \ll 1$
earlier
than in a cosmological constant model with the same matter density and
in addition the relation between the conformal time and angular distance
changes, all of which leads to a larger lensing effect on very small scales.
The value of $\sigma(\theta)/\theta$ linearly increases with $\sigma_8$
in the linear regime, but grows faster than that in the nonlinear
regime. While the value of $\sigma_8$ is still somewhat
uncertain,
it is unlikely that $\sigma_8$ is much bigger
than 1 even in an open model and the curves on figure \ref{fig1}
should indicate the range of the lensing effect in our universe.
To investigate the sensitivity of the
effect to the shape of the power spectrum I compared the
flat model above to the standard CDM model with $\Omega_{m0}=0.5$.
The relative difference between the two models at
$\theta \sim 1'$ only depends on the power spectrum amplitude,
because the dominant scales there are similar to the scales that contribute
to the $\sigma_8$ normalization.
The inverse wavenumber
that makes a dominant contribution to $\sigma(\theta)$
is approximately 0.5 h Mpc for $\theta=1'$ and 0.05 h Mpc at
$\theta=1^\circ$ and above. On larger angular scales the differences
in $\sigma(\theta)/\theta$
between the different spectral shapes increase (with standard CDM
model having less power and thus smaller $\sigma(\theta)/\theta$ for a
given $\sigma_8$ normalization),
but the overall effect is
decreasing and becomes rather small on degree angular scales independent
of the model.

In figure \ref{fig1} the 90 \% c.l. upper
limits on $\sigma(\theta)/\theta$ are also indicated, as
derived from \cite{mould} and
\cite{fahlman} limits on the correlated ellipticities. Both groups
report a null detection of average ellipticity within a
$4.8^\prime$ and $2.76^\prime$
radius aperture, respectively, with a sensitivity of about 1\%.
Adopting median redshifts of $z=0.9$ and $z=0.7$
gives radial distances
$0.27$ and $0.23$ times the comoving distance to the horizon, respectively.
For the two surveys
one obtains upper limits that are comparable to the power spectrum estimates,
which gives additional confidence that the effect was not severely
underestimated.
A general conclusion that can be
derived from these results
is that $\sigma(\theta)/\theta$ is less than 20\%
on scales above $1'$ and less than 5\% on scales larger than $1^\circ$.

\begin{figure}[t]
\vspace*{7.3 cm}
\caption{CMB anisotropy power spectrum $l(l+1)C_l$
versus $l$ with lensing (dashed lines) and
without lensing (solid lines). Upper curves are for adiabatic
CDM model with
$h=0.5$, $\Omega_{m0}=0.4$ and $\Omega_{v0}=0.6$,
lower curves are for adiabatic CDM model with $h=0.5$,
$\Omega_{m0}=1$ and $\Omega_{v0}=0$. Both models are normalized to COBE.
Lensing smoothes the sharp features in
the power spectrum, but leaves the overall shape unchanged. The two
models show a typical range of the lensing effect on CMB.}
\includegraphics{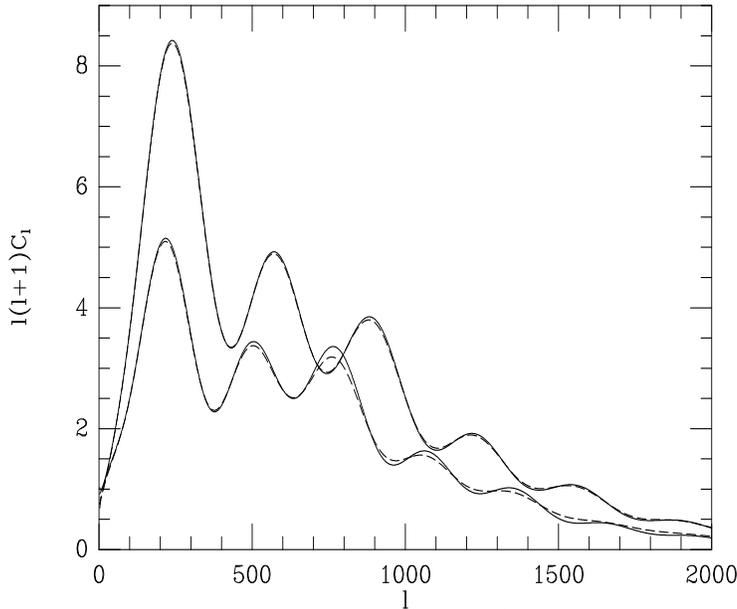}

\label{fig2}
\end{figure}

Figure \ref{fig2} shows the lensing effect on the CMB fluctuation power
spectrum for the models discussed in the previous paragraph.
The CMB multipole moments were obtained from a numerical integration
of perturbed Einstein, Boltzmann and fluid equations (\cite{bode94}).
These models exhibit characteristic acoustic oscillations (Doppler peaks)
and suppression on small scale due to the diffusion
(Silk) damping.
Lensing induces very little gross change
in the power spectrum of CMB. However, the
peaks of acoustic oscillations are smoothed because of lensing and on
smaller angular scales
they can be completely erased. This occurs both because $\sigma(\theta)/\theta$
increases and because the relative width of the
oscillations becomes narrower towards the
smaller angular scales. Observational sensitivity to this effect
depends on the particular experimental setup,
but for most experiments
the window functions are relatively broad in $l$-space and consequently
the effect is diluted.

\begin{figure}[t]
\vspace*{7.3 cm}
\caption{Comparison between several approximations for calculating
the lensing effect on the CMB anisotropies in the COBE
normalized CDM model with $h=0.5$ and $\Omega_{m0}h=0.5$.
Both nonlinear and
linear isotropic
approximations gives results that are almost indistinguishable from
the fully nonlinear and nonisotropic
calculation over this angular range, while
$\epsilon={\rm const}$ approximation gives reliable
results only over a limited range of $l$
and cannot be used for an accurate calculation of the lensing effect.}
\includegraphics{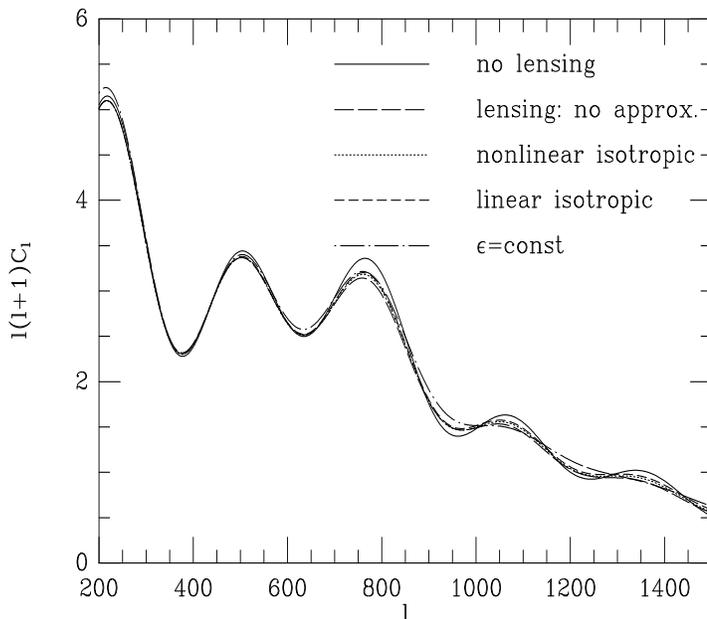}

\label{fig3}
\end{figure}

Currently popular models predict that primary CMB anisotropies are only
important above the Silk damping scale of order of a few arcminutes.
As shown in figure \ref{fig1}, on these scales the linear contributions
to the dispersion $\sigma(\theta)/\theta$ are still dominant and
nonlinear effects are negligible. In fact, using only the linear theory
evolution of power spectrum gives results indistinguishable from
the fully nonlinear calculation in the regime of interest ($l<2000$, see
figure \ref{fig3}).
These angular scales are thus unaffected
by the uncertainties of the nonlinear evolution
and are also the scales where the large-scale observations
place the best constraints on the power spectrum.
The approximation given in equation \ref{clsmooth} gives
reasonable results only if
one adopts $\epsilon$ at $\theta \approx 4\pi l^{-1}$, which is
significantly larger angle than expected (Bond 1995) and again implies
that nonlinear effects are not important until very large $l$.
Even then the agreement is only approximate (figure \ref{fig3}) and
limited to $l<1000$. In general it is better to use the isotropic
approximation in equation
\ref{cthclful} together with the Gauss-Legendre integration to
calculate the lensing
effect on the multipole moments, as it is not significantly
harder to compute than the approximation given in equation \ref{clsmooth}.

\section{Discussion}

The lensing effect on the primary CMB
anisotropies can be calculated
for any cosmological model with a specified evolution of
gravitational potential power spectrum. This
formalism was applied
to several currently popular models,
which best fits the observational data on large scales.
The results indicate that the gravitational lensing does not significantly
affect the CMB power spectrum on degree scales and larger, but becomes
gradually more important towards the smaller scales. Lensing redistributes
the power in the angular correlation function and the amplitude of
the effect depends on the smoothness of the underlying CMB spectrum.
For standard
adiabatic models the acoustic oscillation
peaks are rather prominent even at small
angular scales
(beyond $l\approx 1000$) and lensing may completely erase this structure.

Recently, two groups claimed that the gravitational lensing effect on
CMB has been severly underestimated in previous calculations and that
it importantly changes the CMB pattern even on degree angular scales.
Bassett et al. (1994) assume a model
in which photons propagate through a homogeneous universe with a density
smaller than its mean density to account for the fact that some of the
mass resides in dense clumps. Using the Dyer-Roeder distance-redshift
relation in such a universe they obtain an increase in
angular separation between the two photons relative to its unperturbed value.
Similarly, \cite{fuk} assume a model in which the universe is populated by a
number of massive clumps embedded in a large empty void. Here
the angular separation
between two photons is additionally increased with every passage of the photons
by a clump, because the closer photon is always deflected more
than the more distant one.
This leads to an exponential growth of the angular separation
until it reaches the mean projected separation between the
clumps. One problem with these
models is that they
cannot be applied to the large (supercluster) scales, where the
density fluctuations are small, given that in these models the mass
density in a box on scales smaller than the mean distance between the
clumps is either zero or very large.
Observational data on large scales suggest that density
fluctuations are close to gaussian and
both underdensities and overdensities
have to be included for a proper description of light propagation.
The effect of underdensities is to decrease the angular separation between
the two photons and this leads to a random walk growth of rms deviation
between them. As shown in this paper the lensing effect on CMB on
arcminute scales and larger is dominated by the linear regime, where
underdensities and overdensities play equivalent roles.
Numerical studies of light propagation in realistic models (\cite{jar91})
confirm that there are no large
distortions in the relative photon trajectories present for most lines of
sight,
at least on scales above their resolution scale of a few arcseconds.
Another problem with the above models is that flux conservation
requires that the angular distance
between the photons
(as defined in a homogeneus universe with the same density parameter)
remains on average unchanged, implying that exponential growth
in separation
between photons passing on the same side of a clump is
balanced by a strong focusing of photons that pass on the opposite sides
of the clump. This would lead to multiple images (strong lensing), but it
is known observationally that such situations are rare in our universe,
especially on very large scales, such as superclusters.
It is nevertheless possible that in a highly nonlinear
regime (i.e. on very small scales)
our universe could be well approximated by the models discussed
by \cite{fuk} and Bassett et al. (1994). In such a regime
the models presented in this paper should predict
relative fluctuations larger than unity and their predictions would become
unreliable, because
the assumption that the potential deflecting the photons
can be calculated along the
unperturbed paths
would not be satisfied. As long as $\sigma(\theta)/\theta$ remains small
this is not the case and for the density fluctuations as measured
in our universe this condition is
satisfied at least on angular scales above a few arcseconds.

Although gravitational lensing is of small significance for the present
day experiments, mostly sensitive to the degree angular scales, it may become
relevant for the future experiments that will probe smaller angular
scales with a much higher sensitivity and sky coverage. Gravitational
lensing effect will be especially important for the high precision
determination of cosmological parameters planned for the next generation
of experiments. The uncertainties caused by the gravitational lensing
should be included in the modelling of extraction of cosmological
parameters from the CMB measurements. The formalism developed in this paper
allows to calculate the
lensing effect on the CMB for any specified cosmological
model and can be included as a postprocessor to the standard
calculations of the CMB multipole moments.

\acknowledgments
I would like to thank Ed Bertschinger, Dick Bond
and Enrique Mart\' inez-Gonz\' alez
for useful discussions.
This work was supported by grant NASA NAG5-2816.
\appendix
\section{Appendix}
Calculating the gravitational lensing effect on the CMB
can be cumbersome in general, but it simplifies considerably if
only small angular scales are considered and if the fluctuations in
relative separation between the two photons can be considered
gaussian. The first assumption is not very restrictive, since one does not
expect the lensing effect to be important on large angular scales.
The second assumption should really limit the validity of the
calculation to the linear scales only, where
the prediction of most models that the initial fluctuations are gaussian
guarantees its validity. In reality its validity extends beyond that
to the quasi-linear scales, because the relative fluctuations are
obtained by a projection of a 3-dimensional distribution over a
broad radial window function and are in general much more gaussian than the
3-d distribution of the gravitational potential derivative itself.

In the spherical plane approximation
one can write the temperature anistropies ${\Delta T \over T}(\vec \theta)$
in terms of its Fourier transform,
\begin{equation}
{\Delta T \over T}(\vec \theta)=\int d^2\vec l e^{-i\vec l \cdot \vec
\theta} {\Delta T \over T} (\vec l).
\end{equation}
The correlation function including lensing is given by
\begin{eqnarray}
\tilde{C}(\theta)=\langle {\Delta T \over T}(\vec \theta^A+\delta \vec
\theta^A){\Delta T\over T}(\vec \theta^B +\delta \vec
\theta^B) \rangle_{\vec \theta^A \cdot \vec \theta^B=\cos \theta}=
\nonumber \\
=\int d^2\vec l \int d^2\vec l' e^{-i\vec l \cdot \vec
\theta^A+i\vec l' \cdot \vec \theta^B} \biggl\langle
e^{-i\vec l \cdot \delta \vec
\theta^A+ i\vec l' \cdot \delta \vec \theta^B}
{\Delta T \over T}(\vec l) {\Delta T \over T}^*(\vec l') \biggl\rangle,
\label{ctil}
\end{eqnarray}
where $\delta \vec
\theta^A$ and $\delta \vec
\theta^B$ are the angular excursions of the two photons that are separated
at the observer's position by an angle $\theta$.
In equation \ref{ctil} one has to average both over the intrinsic temperature
anisotropies ${\Delta T \over T}(\vec l)$
and over the lensing fluctuations $\delta \vec\theta$. The first
averaging gives the angular power spectrum of CMB,
\begin{equation}
\biggl\langle {\Delta T \over T}(\vec l) {\Delta T \over T}^*(\vec l')
\biggl\rangle=
C_l {\delta^2(\vec l -\vec l') \over (2\pi)^2},
\end{equation}
while the second gives the characteristic function of a gaussian
field $\delta \vec\theta^A-\delta \vec\theta^B$,
\begin{equation}
\langle e^{-i\vec l \cdot(\delta \vec
\theta^A-\delta \vec
\theta^B)}\rangle=
e^{-<[\vec l\cdot(\delta \vec\theta^A-\delta \vec\theta^B)]^2>/2}.
\end{equation}
Performing the ensemble averaging over the lensing fluctuations $\delta
\vec\theta$
in the equation above with the help of
Limber's equation (Kaiser 1992) leads to the
lensed correlation function
\begin{equation}
\tilde{C}(\theta)=(2 \pi)^{-2}\int_0^\infty l\,dl\,C_l
\int_0^{2\pi}d\varphi_l\,\exp\left[{-l^2 \over 2}
[\sigma^2(\theta)-\cos(2\varphi_l)C_{\rm
gl,2}(\theta)]-il\theta\cos(\varphi_l)\right],
\label{cful}
\end{equation}
where $\sigma(\theta)$ is the rms dispersion in the
angular positions of the two photons defined in equation \ref{limber}
and $C_{\rm gl,2}(\theta)$ can be obtained from $C_{\rm gl}(\theta)$
in equation \ref{limber} by replacing $J_0$ with $J_2$. Assuming
$l^2C_{\rm gl,2}(\theta)\ll 1$ (which follows from
assuming $\sigma(\theta)/\theta \ll 1$,
together with $l \sim \theta^{-1}$
and $C_{\rm gl,2}(\theta) <
\sigma^2(\theta)$)
one may Taylor expand the exponential in equation \ref{cful}
and integrate over $\varphi_l$.
This leads to
\begin{equation}
\tilde{C}(\theta)=(2 \pi)^{-1}\int_0^\infty l\,dl e^{-\sigma^2
(\theta)l^2/ 2} C_l \left[
J_0(l\theta)+{l^2 \over 2}C_{\rm gl,2}(\theta)J_2(l\theta)\right].
\label{cthclful}
\end{equation}

One
approximation often used in the literature is to keep only the
isotropic term $J_0(l\theta)$ in equation \ref{cthclful},
which gives the dominant contribution to the lensing
effect. In the following I will assume this approximation, because it
gives a good agreement with the exact calculation,
as shown in figure \ref{fig3}.
With this approximation the lensed correlation function becomes
\begin{equation}
\tilde{C}(\theta)=(2 \pi)^{-1}\int_0^\infty l\,dl e^{-\sigma^2
(\theta)l^2/ 2} C_l J_0(l\theta).
\label{cthcl}
\end{equation}
This equation is essentially the same as the Wilson \&
Silk (1981) expression for the correlation function
observed with an instrument that has a gaussian beam profile, the
only difference being that in the present case the dispersion
$\sigma(\theta)$ depends on the angular separation $\theta$.
After another Fourier transform and using equation 6.615 from
Gradshteyn \& Ryzhik (1965)
one obtains equation \ref{c}.
Alternatively, one can also express the lensing effect directly in
terms of the CMB power spectrum,
\begin{equation}
\tilde{C}_l=\int_0^\pi \theta d\theta \int_0^\infty l'\,dl' e^{-\sigma^2
(\theta)l'^2 /2} C_{l'} J_0(l\theta) J_0(l'\theta).
\label{tilcl}
\end{equation}

The above expression can be further simplified if one assumes that
$\epsilon=\sigma(\theta)/\theta$ is a constant.
Equation \ref{tilcl} then reduces to
\begin{equation}
\tilde{C}_l=\int_0^\infty {l' \, dl' \over (\epsilon l')^2}
C_{l'} I_0\left({l \over \epsilon^2 l'} \right)\exp\left[{l^2+l'^2 \over
2(\epsilon l')^2}\right],
\label{cleps}
\end{equation}
where $I_0(x)$ is the modified Bessel function of order 0. Further
assuming $\epsilon \ll 1$ one can asymptotically expand the modified
Bessel function $I_0(x)$ and use $l' \approx l$ everywhere except in
the exponential. This finally leads to equation \ref{clsmooth}, which is
similar to the expression given by Bond (1995).

\end{document}